\documentstyle[11pt,epsfig,amssymb,amsmath]{article}
\title{\bf  QCD phase transition in DGP brane cosmology}
\author{K. Atazadeh$^1$\thanks{email: atazadeh@azaruniv.edu},\, A. M. Ghezelbash$^2$\thanks{masoud.ghezelbash@usask.ca} \,\
and H. R. Sepangi$^3$\thanks{email: hr-sepangi@sbu.ac.ir}
\\{\small $^1$Department of Physics, Azarbaijan University of Tarbiat Moallem, Tabriz 53741-161, Iran} \\ {\small $^2$Department of Physics and
Engineering Physics, University of Saskatchewan,}\\{\small
Saskatoon, Saskatchewan S7N 5E2, Canada}\\{\small $^3$Department of Physics, Shahid Beheshti University, Evin,
Tehran 19839, Iran}}
\begin{document}
\maketitle
\begin{abstract}
In the standard picture of cosmology it is predicted that a phase transition,
associated with chiral symmetry breaking after the electroweak
transition, has occurred at approximately $10 \mu$ seconds after the
Big Bang to convert a plasma of free quarks and gluons into hadrons.
We consider the quark–-hadron phase transition in a
DGP brane world scenario within an effective model of QCD. We study
the evolution of the physical quantities useful for the study of
the early universe, namely, the energy density,
temperature and the scale factor before, during, and after the phase
transition. Also, due to the high energy density in the early
universe, we consider the quadratic energy density term that appears
in the Friedmann equation. In DGP brane models such a term
corresponds to the negative branch ($\epsilon=-1$) of the Friedmann
equation when the Hubble radius is much smaller than the crossover
length in $4D$ and $5D$ regimes. We show that for different values
of  the cosmological constant on a brane, $\lambda$, phase transition occurs and  results in
decreasing the effective temperature of the quark--gluon plasma and
of the hadronic fluid. We then
consider the quark–hadron transition in the smooth crossover regime at high and low temperatures
and show that such a transition occurs along with decreasing the effective temperature of the
quark-gluon plasma during the process of the phase transition.
\end{abstract}

\section{Introduction}

Over the recent past,  brane-world \cite{Rubakov} scenarios have
generated a great deal of interest.
This interest was motivated by the fact that there is a strongly coupled
sector of $E_8 \times E_8$ heterotic string theory which can be
described by a field theory living in an 11-dimensional space-time
\cite{horava}. The 11-dimensional world consists of two
10-dimensional hypersurfaces embedded on the fixed points of an
orbifold and the matter fields are assumed to be confined and live
on these hypersurfaces which are known to be 9-branes. After
compactification of the 11-dimensional theory on a Calabi-Yau
3-fold, we obtain an effective 5-dimensional theory \cite{lukas}
which has the structure of two 3-branes located on the orbifold
boundaries. This scenario has motivated intense efforts to
understand the case where the bulk is a 5-dimensional anti
de-Sitter space. In this setup, gravitons are allowed to penetrate
into the bulk but are localized on and around the brane
\cite{lisa}. It was then shown that in a background of a
non-factorizable geometry an exponential warp factor emerges which
multiplies the Poincar\'e invariant 3+1 dimensions in the metric.
The model consists of two $3$-branes situated along the 5th
dimension, compactified on a $S^{1}/Z_2$ orbifold symmetry where
the two branes must have opposite tensions. The evolution equation
followed from such a brane scenario differs from  that of the
standard four dimensional evolution equation when no branes are
present~\cite{bin}. The existence of branes and the requirement
that matter fields should be localized on the brane lead to a
non-conventional cosmology, necessitating a more concerted study.
A large number of studies have been devoted to the effective
gravity induced on the brane~\cite{brane} and, in particular, a
great amount of interest has been generated regarding inflationary
cosmology~\cite{numero}. Not surprisingly, the problem of the
cosmological constant has become a focal point in the brane-world
studies where, for example, in \cite{nim,kachru,bin1} a five
dimensional action with a scalar field is non-minimally coupled to
five dimensional gravity and to the four dimensional brane
tension. There has also been some discussion on the localization
of gravity \cite{gomez}. A feature common to these type of models
is that they predict deviations from the usual $4D$ gravity at
short distances.

A somewhat different approach within the brane-world framework is
the model proposed by Dvali, Gabadadze and Porrati (DGP)
\cite{dvali,dvali2}. It predicts deviations from the standard $4D$
gravity over large distances. The transition between four and
higher-dimensional gravitational potentials in  DGP model arises
because of the presence of both the brane and bulk Einstein terms
in the action. The Friedmann-like equations governing the
cosmological evolution of a brane possessing an intrinsic
curvature term in the action have already been derived and
discussed for an AdS-Schwarzschild bulk space-time \cite{holdom}.
Cosmological consideration of the DGP model was first discussed in
\cite{deffayet} where it was shown that in a Minkowski bulk
space-time self-accelerating solutions exist. In the
original DGP model it is known that $4D$ general relativity is not
recovered at linearized level. However, some authors have shown
that at short distances we can recover the $4D$ general relativity
in a spherically symmetric configuration, see for example
\cite{tanaka}. An important observation was made in
\cite{dick,dick2} where it was shown that DGP model allows for an
embedding of the standard Friedmann cosmology whereby the
cosmological evolution of the background metric on the brane can
entirely be described by the standard Friedmann equation plus
energy conservation on the brane. This was later extended to
arbitrary number of transverse dimensions in \cite{Cordero}. For a
comprehensive review of the phenomenology of DGP cosmology, the
reader is referred to \cite{Lue}.

Standard cosmology suggests that as the early universe expanded
and cooled, it underwent a series of symmetry-breaking phase
transitions, causing topological defects to form. It is the study
of such phase transitions that would pave the way for a better
understanding of the evolution of the early universe,
characterized by the existence of a quark-gluon plasma undergoing
a phase transition. In what follows we focus attention on possible
scenarios which might have occurred to allow the phase transition
mentioned above come to the fore. We generally follow the
discussion presented in \cite{harko,heydari,ata} which puts the quark-gluon
phase transition in a cosmologically transparent perspective.

The existence of the phase transition from the quark–-gluon plasma phase to hadron gas phase is therefore a
definite prediction of QCD. However, the phase transition in QCD can be characterized by a truly
singular behavior of the partition function leading to a first or second order
phase transition, or it can be only a crossover with rapid changes in
some observables, strongly depending on the values of the quark masses.
The possibility of phase transitions in a gas of quark–-gluon bag was
demonstrated for the first time in \cite{Gorenstein}. Most studies have shown that first, second or
higher order transitions are possible. The possibility of no phase transitions has also been pointed out in \cite{Greiner}.
Recently, lattice QCD calculations for two quark flavors suggest
that QCD makes a smooth crossover transition at a
temperature of $T_c\sim 150 $ MeV \cite{tan}. Such a phase
transition could be responsible for the formation of relic
quark--gluon objects in the early universe which may have survived.
In this paper, our studies on phase transition are based on the ideas proposed in the first reference in \cite{Gorenstein},
where it was shown that under certain circumstances a gas of extended
hadrons could produce phase transitions of the first or second
order, and also a smooth crossover transition that might be qualitatively similar to that of lattice QCD. The physics of  quark–-hadron phase transition and its
cosmological implications have been extensively discussed in the
framework of general relativistic cosmology in \cite{quark}-\cite{quark12}.

As is well known, the Friedmann equation in brane-world scenarios
differs from that of the standard $4D$ cosmology which results in
an increased expansion rate in early times. We expect this
deviation from the standard $4D$ cosmology to have noticeable
effects on the cosmological evolution, especially on cosmological
phase transitions. In the context of brane-world models, the first
order phase transitions have been studied in \cite{DaVe01} where
it was shown that due to the effects coming from higher
dimensions, a phase transition requires a higher nucleation rate
to complete, and, baryogenesis and particle abundances could be
suppressed. Recently, the quark-hadron phase transition was
studied  in a Randall-Sundrum brane-world scenario \cite{harko}.
Within the framework of phase transitions, the authors
studied the evolution of the relevant cosmological parameters
(energy density, temperature, scale factor, etc.) of the
quark-gluon and hadron phases and the phase transition itself. It
would therefore of interest to study first order and crossover phase transitions in
the context of a $\lambda$DGP brane-world scenario and this is what we
intend to do in what follows.

\section{Field equations in a DGP brane scenario}
We start by writing the action for a $\lambda$DGP brane-world
\begin{equation}\label{eq2.1}
{\cal S}=\frac{M_{5}^{3}}{2}\int_{{\rm Bulk}} d^{5}x\sqrt{-g}{\cal R}+
\int_{{\rm brane}} d^{4}x\sqrt{-q} \left(\frac{M_{4}^{2}}{2}R-\lambda\right) +{\cal S}_{m}[q_{\mu\nu},\psi_m],
\end{equation}
where the first term corresponds to the Einstein–-Hilbert action in
$5D$ with the $5$-dimensional bulk metric $g_{_{AB}}$ and the Ricci
scalar  ${\cal R}$. Similarly, the second term is the
Einstein–-Hilbert action corresponding to the induced metric
$q_{\mu\nu}$ on the brane, where $R$ is the relevant scalar
curvature and $\lambda$ is cosmological constant on the brane. Also, $M_4=(8\pi G_{4})^{-\frac{1}{2}}$ and $M_5=(8\pi
G_{5})^{-\frac{1}{3}} $ are the reduced Planck masses in four and five
dimensions respectively and ${\cal S}_m$ is the matter action on the
brane with matter field $\psi_m$. The induced metric $q_{\mu\nu}$ is
defined as usual from the bulk metric $g_{_{AB}}$ by
\begin{equation}\label{eq2.2}
q_{\mu\nu} = \delta^{A}_{\mu} \delta^{B}_{\nu }g_{_{AB}}.
\end{equation}
Variation of the action with respect to $g_{_{AB}}$ yields the
field equations
\begin{equation}\label{eq2.3}
G_{AB}+ (2 \ell_{_{\it DGP}}\delta_{A}^{\mu}\delta_{B}^{\nu}G_{\mu\nu}-\lambda\delta_{A}^{\mu}\delta_{B}^{\nu}q_{\mu\nu})\delta(y)=T_{AB},
\end{equation}
where $T_{AB}$ is the  energy-momentum tensor of the matter content
of the $5$-dimensional space-time and $\ell_{_{\rm
DGP}}=\frac{M_4^{2}}{2M_5^{3}}$ is the crossover length scale
between the $4D$ and $5D$ regimes in the DGP brane model. Note that
the brane is the hypersurface defined at $y=0$. To continue, we
follow the approach presented in \cite{deffayet} where it was
first shown how one recovers a self-accelerating solution from the
DGP field equations (\ref{eq2.3}).

Being interested in cosmological solutions, we take a metric of the form
\begin{equation} \label{eq2.4}
ds^{2} = -n^{2}(\tau,y) d\tau^{2}
         +a^{2}(\tau,y)\delta_{ij}dx^{i}dx^{j}
         +b^{2}(\tau,y)dy^{2}\ ,
\end{equation}
where $\gamma_{ij}$ is a maximally symmetric 3-dimensional metric
$(k=-1,0,1$ will parameterize the spatial curvature) and we focus
our attention on a spatially flat ($k=0$) space-time. Since matter
resides on the brane and the bulk is taken to be empty, the field
equations (\ref{eq2.3}) in the bulk can be written as
\begin{eqnarray}
    G_0^0 &=& {3\over n^2}\left[{\dot{a}\over a}\left({\dot{a}\over a} + {\dot b\over b}\right)\right]
        - {3\over b^2}\left[{a''\over a} + {a'\over a}\left({a'\over a} - {b'\over b}\right)\right] = 0~,  \\
    G_j^i &=& {1\over n^2}\delta_j^i\left[-{2\ddot a\over a} - {\ddot b\over b}
            + {\dot a\over a}\left({2\dot n\over n} - {\dot a\over a}\right)    \nonumber
            + {\dot b\over b}\left({\dot n\over n} - {2\dot a\over a}\right)\right]  \\
        && \ \ \ + {1\over b^2}\delta_j^i\left[{n''\over b} +{2a''\over a}
            + {a'\over a}\left({2n'\over n} + {a'\over a}\right)
            - {b'\over b}\left({n'\over n} +{2a'\over a}\right)\right] = 0~,  \\
    G_5^5 &=& -{3\over n^2}\left[{\ddot a\over a}
        + {\dot{a}\over a}\left({\dot{n}\over n} + {\dot a\over a}\right)\right]
        - {3\over b^2}\left[{a'\over a}\left({n'\over n} + {a'\over a}\right)\right] = 0~,  \\
    G_{05} &=& 3\left[-{\dot a'\over a} + {\dot a\over a}{n'\over n} + {\dot b\over b}{a'\over a}\right] = 0\ ,
\end{eqnarray}
where a prime denotes differentiation with respect to $y$ and a dot
denotes differentiation with respect to $\tau$. By using a technique
first developed in ~\cite{bin,Flanagan:1999cu}, the bulk equations,
remarkably, may be solved exactly given that the bulk is devoid of
matter. Assuming $\textit{Z}_2$ symmetry across the brane, we find
that the metric components in equation ~(\ref{eq2.4}) are given by
\cite{deffayet}
\begin{eqnarray}\label{eq2.5}
n(\tau, y) &=& 1 \mp |y| {\ddot{a_{_{0}}}\over \dot{a_{_{0}}}}~, \nonumber \\
a(\tau, y) &=& a_{_{0}} \mp |y| \dot{a_{_{0}}}~,     \\
b(\tau, y) &=& 1\ , \nonumber
\end{eqnarray}
where there remains a parameter $a_{_{0}}(\tau)$ that is yet to be
determined. Note that $a_0(\tau) = a(\tau, y=0)$ represents the
usual scale factor of the four-dimensional cosmology in our brane
universe. According to equations (\ref{eq2.5}), there are  two
distinct possible cosmologies associated with the choice of the sign.

To obtain cosmological equations we must have the total
energy-momentum tensor which includes matter and the cosmological
constant on the brane which may be written as
\begin{equation}\label{eq2.6}
T^A_B|_{\rm brane}= ~\delta (y)\ {\rm diag}
\left(-\rho,p,p,p,0 \right)\ .
\end{equation}
In order to determine the scale factor $a_{_{0}}(\tau)$, we need to
employ the proper boundary condition on the brane.  This can be done
by taking equations (\ref{eq2.3}) and integrating across the brane.
Then, the boundary conditions require
\begin{eqnarray}\label{eq2.7}
    \left.{n'\over n}\right|_{y=0} &=& - {8\pi G_4\over 3}\rho, \\
    \left.{a'\over a}\right|_{y=0} &=& {8\pi G_4\over 3}(3p + 2\rho)\ .
\end{eqnarray}
Comparison with the bulk solutions (\ref{eq2.5}) imposes a
constraint on the evolution of $a_{_{0}}(\tau)$.  Such an
evolution is tantamount to a new set of Friedmann equations
\cite{deffayet}
\begin{eqnarray}\label{eq2.8}
H^2 -{\epsilon\over \ell_{_{\rm DGP}}}H = {8\pi G_4\over 3}\rho(\tau)+\lambda\ .
\end{eqnarray}
and
\begin{eqnarray}\label{eq2.9}
\dot{\rho} + 3(\rho+p)H = 0\ ,
\end{eqnarray}
where $\epsilon=\pm1$ and we have used the usual Hubble parameter $H
= \dot a_{_{0}}/a_{_{0}}$.  The second of these equations is just
the usual expression of energy-momentum conservation.  The first
equation, however, is indeed a new Friedmann equation that is a
modification of the ordinary four-dimensional Friedmann equation with a cosmological constant on brane.
There are several constraints obtained for the brane cosmological constant $\lambda$. Thus from the big bang nucleosynthesis constraint it follows that $\lambda \geq1$ MeV$^{4}$ \cite{maartens}. A much stronger bound for $\lambda$
arises from null results of submillimeter tests of Newton's law, giving $\lambda \geq10^{8}$ GeV$^{4}$ \cite{maartens2}. An
astrophysical lower limit on $\lambda$, which is independent of the Newton law and the cosmological
limits has been derived in \cite{maartens}, leading to $\lambda>5\times10^{8}$ MeV$^4$.

Equations (\ref{eq2.8}) and (\ref{eq2.9}) together with the brane
equation of state are sufficient to derive the cosmological
evolution of the brane metric. Now, assuming $\rho\geq0$, equation
(\ref{eq2.8}) can be rewritten as
\begin{eqnarray}\label{eq2.10}
    H={\epsilon\over 2\ell_{_{\rm DGP}}}+\sqrt{\frac{1}{4\ell_{_{\rm DGP}}^2}+{8\pi G_4\over 3}\rho+\lambda} .
\end{eqnarray}
Let us consider equation (\ref{eq2.8}) more closely. The new
contribution from the DGP brane-world  is the appearance
of the term ${\epsilon\over \ell_{_{\rm DGP}}}H$ on the left-hand
side of the Friedmann equation. The choice for the sign of
$\epsilon$ represents two distinct cosmological phases. Just as
gravity is modified at large scales and behaves $4D$-like at short
distances, so too is the Hubble scale, $H(\tau)$, which evolves by
the conventional Friedmann equation at high Hubble scales but is
altered substantially as $H(\tau)$ approaches $\ell_{_{\rm
DGP}}^{-1}$.  There is therefore the possibility of two distinct
cosmological phases \cite{deffayet}. First,  the phase in equation
(\ref{eq2.8}) corresponding to the choice  $\epsilon=-1$ which had
already been studied in
~\cite{holdom,Shtanov:2000vr,Nojiri:2000gv}, with transitions from
$H^2\sim \rho$ to $H^2\sim \rho^2$. We refer to this phase as the
Friedmann-–Robertson–-Walker (FRW) phase. The other cosmological
phase corresponds to $\epsilon=1$. Here, cosmology at early times
again behaves according to the conventional four-dimensional
Friedmann equation, but at late times approaches a brane
self-accelerating phase.
\section{Quark--hadron phase transition}
The quark-hadron phase transition is a notion fundamental to the
study of particle physics, particularly in the context of lattice
gauge theories. However, it has become increasingly relevant and
even an integral part of any study dealing with the underlying
mechanisms responsible for the evolving universe at its early stages
of formation in which a soup of quarks and gluons interact and
assume to undergo a first order phase transition to form hadrons. It is
therefore essential to have a review of the basic ideas before
attempting to use the results obtained from such a phase transition
and apply them to the study of the evolution of the early universe
within the context of a DGP brane-world scenario. In this regard, a
well written and concise review can be found in \cite{harko} and the
interested reader should consult it. Here, it would suffice to
mention the results relevant to our study and leave the details of
the discussion to the said reference.

We start from the equation of state of matter in the quark phase
which can generally be given in the form
\begin{equation}\label{eq3.1}
\rho_q=3a_qT^{4}+V(T),~~~~~~~~~~~~p_q=a_qT^{4}-V(T),
\end{equation}
where $a_q = (\pi^{2}/90)g_q $, with $g_q = 16 + (21/2)N_F + 14.25
= 51.25$ and $N_F = 2$ with $ V (T )$ being the self-interaction
potential. For $V(T)$ we adopt the expression \cite{quark10}
\begin{equation}\label{eq3.2}
V(T)=B+\gamma_T T^{2}-\alpha_{T}T^{4},
\end{equation}
where $B$ is the bag pressure constant,  $\alpha_T = 7\pi ^{2}/20$
and $\gamma_T = m^{2}_s /4$ with $m_s$ the mass of the strange
quark in the range $m_s \in(60–-200)$ MeV. In the case where the
temperature effects are ignorable, the equation of state in the
quark phase takes the form of the MIT bag model equation of state,
$p_q = (\rho_q- 4B)/3$. Results obtained in low energy hadron
spectroscopy, heavy ion collisions and phenomenological fits of
light hadron properties give $B^{1/4}$ between $100$ and $200$ MeV
\cite{LePa92}.

Once the hadron phase is reached,  one takes the cosmological fluid
with energy density $\rho_h$ and pressure $p_h$ as an ideal gas of
massless pions and nucleons obeying the Maxwell-–Boltzmann
statistics. The equation of state can be approximated by
\begin{equation}\label{eq3.3}
p_h(T)=\frac{1}{3}\rho_h(T)=a_\pi T^4,
\end{equation}
where $a_\pi =(\pi^2/90)g_h$ and $g_h=17.25$.
The critical temperature $T_c$ is defined by the condition
$p_q(T_c) = p_h(T_c)$ \cite{quark}, and is given by
\begin{equation}\label{eq3.4}
T_c=\left[\frac{\gamma_T+\sqrt{\gamma_T^{2}+4B(a_q+\alpha_T-a_\pi)}}{2(a_q+\alpha_T-a_\pi)}\right]^{1/2}.
\end{equation}
If we take  $m_s = 200$ MeV and $B^{1/4} = 200$ MeV,  the
transition temperature is of the order $T_c \approx 125$ MeV. It
is worth mentioning that since the phase transition is assumed to
be of first order, all the physical quantities exhibit
discontinuities across the critical curve.

\section{Dynamical consequences of DGP brane universe during quark--hadron phase transition}
We are now in a position to study the phase transition described
above. The framework we are working in is defined by the DGP
brane-world scenario for which the basic equations were derived in
section 2 and therefore the essential physical quantities that
should be monitored through the quark-hadron phase transition are
the energy density $\rho $, temperature $T$ and scale factor
$a_{_{0}}$. These parameters are determined by the Friedmann
equation (\ref{eq2.8}), conservation equation (\ref{eq2.9}) and
the equations of state, namely (\ref{eq3.1}), (\ref{eq3.2}) and
(\ref{eq3.3}). We should now consider the evolution of the DGP
brane-world before, during and after the phase transition era.

Let us consider the era preceding the phase transition for which
$T>T_{c}$ and the universe is in the quark phase. Use of
equations of state of the quark matter and the conservation of
matter on the brane, equation (\ref{eq2.9}),  leads to
\begin{equation}\label{eq4.1}
H=\frac{\dot{a_{_{0}}}}{a_{_{0}}}=-\frac{3a_{q}-\alpha
_{_{T}}}{3a_{q}}\frac{\dot{T}}{T}-\frac{1}{6}\frac{\gamma
_{_{T}}}{a_{q}}\frac{\dot{T}}{T^{3}}.
\end{equation}
Integrating the above equation immediately gives
\begin{equation}\label{eq4.2}
 a_{_{0}}(T)=c~T^{\frac{\alpha _{_{T}}-3a_{q}}{3a_{q}}}\exp \left(
\frac{1}{12}\frac{\gamma _{_{T}}}{a_{q}}\frac{1}{T^{2}}\right) ,
\end{equation}
where $c$ is a constant of integration. We may now proceed to obtain an expression describing the
evolution of temperature of the DGP brane universe in the quark
phase by combining  equations (\ref{eq4.1}) and (\ref{eq2.8}),
giving
\begin{equation}\label{eq4.3}
\frac{dT}{d\tau}=\frac{-1}{A_{_{0}} T^2+A_{_{1}}}\left(\frac{\epsilon~ T^3}{2\ell_{_{\rm DGP}}}+\sqrt{
\frac{8\pi G_4}{3}\left(a_q-\frac{\alpha_{_{T}}}{3}\right)T^{10}+\frac{8\pi G_4}{3}
\gamma _{_{T}}T^{8}+\left(\frac{1}{4\ell_{_{\rm DGP}}^2}+\frac{8\pi G_4}{3}B+\lambda\right)T^{6}}\right),
\end{equation}
where we have denoted
\begin{eqnarray}\label{eq4.4}
A_{0}=1-\frac{\alpha_{_{T}}}{3a_{_{T}}},
\\\nonumber
A_{1}=\frac{\gamma_{_{T}}}{6a_{q}}.
\end{eqnarray}
Equation (\ref{eq4.3}) may now be solved numerically and the
result is presented in Figure 1 which shows the behavior of
temperature as a function of cosmic time $\tau$ in a DGP
brane-world filled with quark matter for different values of
$\lambda$.
\subsection{Constant self-interaction potential $(V(T)=B)$}
One may gain an analytical understanding of the evolution of
cosmological quark matter in our DGP brane-world by taking the
negative branch ($\epsilon=-1$) in  equation (\ref{eq2.8}). Let us
concentrate on the phase transition era, namely
$H^{-1}\ll\ell_{_{\rm DGP}}$ and take the simple case where
temperature corrections can be neglected in the self-interaction
potential $V$. Then $V=B={\rm cons.}$ and equation of state of the
quark matter is given by that of the bag model, $p_{q}=\left( \rho
_{q}-4B\right) /3$. Equation (\ref{eq2.9}) may then be integrated
to give the scale factor on the brane as a function of temperature
\begin{equation}\label{eq4.5}
a_{_{0}}(T)=\frac{c}{T},
\end{equation}
where $c$ is a constant of integration.

Since in the early universe the density of matter is very high and
the Hubble radius ($H^{-1}$) is  small with respect to the
crossover length scale $(\ell_{_{\rm DGP}})$ {\it i.e.}
$H^{-1}\ll\ell_{_{\rm DGP}}$~, equation (\ref{eq2.8}) for
$\epsilon=-1$  leads to  $\rho ^{2}$ {\it i.e.} $H^{2}\sim\rho
^{2}$ which is the specific character of brane world models. The
evolution of the quark phase of the DGP brane-world is now
described by the equation
\begin{equation}\label{eq4.6}
\frac{\dot{a_{_{0}}}}{a_{_{0}}}=\frac{8\pi G_{4}}{3} \rho +\lambda.
\end{equation}
\begin{figure}
\begin{center}
\epsfig{figure=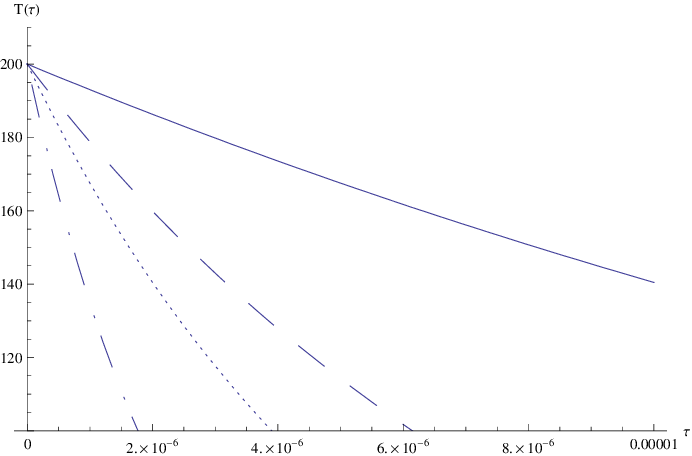,width=8cm}
\end{center}
\caption{\footnotesize  The behavior of $T(\tau)$ as a function of
time ($\tau$) for the positive branch ($\epsilon=1$) and different
values of $\lambda$: $\lambda= 8\times10^{8}$
MeV$^{4}$ (solid curve), $\lambda=8\times 10^{9}$
MeV$^{4}$ (dashed curve), $\lambda=2\times 10^{10}$
MeV$^{4}$ (dotted curve) and $\lambda=1\times 10^{11}$
MeV$^{4}$ (dotted-dashed curve). We have taken $B^{1/4}=200$ MeV
, $G_4\sim10^{-44}$ MeV$^{-2}$ and $\ell_{_{\rm DGP}}\sim10^{-38}$ MeV$^{-1}$ \cite{dvali}.}
\end{figure}
Using equations (\ref{eq2.9}) and (\ref{eq4.6}), the time
dependence of temperature can be obtained from  equation
\begin{equation}\label{eq4.7}
\frac{dT}{d\tau}= \frac{-8\pi G_{4}}{3} \left(
3a_{q}T^{5}+BT\right)-\lambda T.
\end{equation}
This equation has the following general solution
\begin{equation}\label{eq4.8}
T(\tau)= \frac{(-\frac{8\pi G_{4}}{3}B+\lambda)^{1/4}}{\left[ 8\pi G_{4}a_{q}-C\exp
 (\frac{32\pi G_{4}B}{3}-4\lambda)\tau \right] ^{1/4}}.
\end{equation}
The integration constant $C$ can be obtained by relating it to the
temperature $T|_{\tau=\tau_0}\equiv T_0$ of the quark matter at the
time $\tau_{0}$ using the relation
\begin{equation}\label{eq4.9}
C=\frac{8\pi G_{4}a_{q}T_0^{4}+\frac{8\pi G_{4}}{3}B+\lambda}{\exp(\frac{32\pi G_{4}B}{3}-4\lambda)\tau_{0}}.
\end{equation}
\subsection{Formation of hadrons}
During the phase transition, temperature and pressure are
constant and quantities like the  entropy $S=sa^{3}$ and enthalpy
$W=\left( \rho +p\right) a^{3}$ are conserved. Following
\cite{quark,harko}, we replace $\rho \left( \tau\right) $ by
$h(\tau)$, the volume fraction of matter in the hadron phase, by
defining
\begin{equation}\label{eq4.11}
\rho \left( \tau\right) =\rho _{_{H}}h(\tau)+\rho _{_{Q}}\left[ 1-h(\tau)\right]
=\rho _{_{Q}} \left[ 1+nh(\tau)\right] ,
\end{equation}
where $n=\left( \rho _{_{H}}-\rho _{_{Q}}\right) /\rho _{_{Q}}$.
The beginning of the phase transition is characterized by
$h(\tau_{c})=0$ where $\tau_{c}$ is the time representing the
beginning of the phase transition and $\rho \left( \tau_{c}\right)
\equiv \rho _{_{Q}}$, while the end of the transition is
characterized by $h\left( \tau_{h}\right) =1$ with $\tau_{h}$
being the time signalling the end and corresponding to $\rho \left(
\tau_{h}\right) \equiv \rho _{_{H}}$. For $\tau>\tau_{h}$ the
universe enters into the hadronic phase.

Equation (\ref{eq2.9}) now gives
\begin{equation}\label{eq4.12}
\frac{\dot{a_{_{0}}}}{a_{_{0}}}=-\frac{1}{3}\frac{\left( \rho _{_{H}}-\rho
_{_{Q}}\right) \dot{h} }{\rho _{_{Q}}+p_{c}+\left( \rho _{_{H}}-\rho
_{_{Q}}\right) h}=-\frac{1}{3}\frac{r \dot{h}}{1+rh},
\end{equation}
where we have denoted $r=\left( \rho _{_{H}}-\rho _{_{Q}}\right)
/\left( \rho _{_{Q}}+p_{c}\right)$. From the above equation we can
obtain the relation between the scale factor on the brane and the
hadron fraction $h(\tau)$
\begin{equation}\label{eq4.13}
a_{_{0}}(\tau)=a_{_{0}}\left( \tau_{c}\right) \left[ 1+rh(\tau)\right] ^{-1/3},
\end{equation}
where the initial condition $h\left(\tau_{c}\right)=0$ has been
used. Now, using equations (\ref{eq2.8}) and (\ref{eq4.13}) we
obtain the time evolution of the matter fraction in the hadronic
phase
\begin{equation}\label{eq4.14}
\frac{dh}{d\tau}=-3\left( h+\frac{1}{r}\right)
\left(\frac{\epsilon}{2\ell_{_{\rm DGP}}}+
\sqrt{\frac{1}{4\ell_{_{\rm DGP}}^2}+\lambda+\frac{8\pi
G_4}{3}\rho_{_{Q}}[1+nh(\tau)]}\right).
\end{equation}
Figure 2 shows variation of the hadron fraction $h(\tau)$ as a
function of $\tau$  for different values of $\lambda$
\begin{figure}
\begin{center}
\epsfig{figure=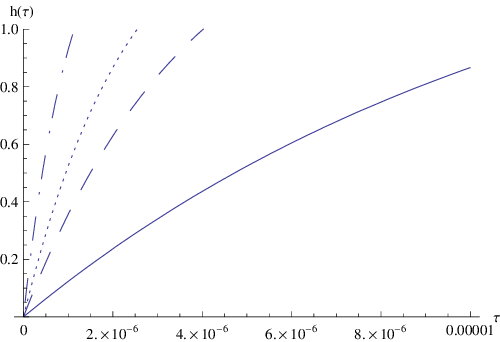,width=8cm}
\end{center}
\caption{\footnotesize  The behavior of $h(\tau)$ as a function of
time ($\tau$) for the positive branch ($\epsilon=1$) and different
values of $\lambda$: $\lambda= 8\times10^{8}$
MeV$^{4}$ (solid curve), $\lambda=8\times 10^{9}$
MeV$^{4}$ (dashed curve), $\lambda=2\times 10^{10}$
MeV$^{4}$ (dotted curve) and $\lambda=1\times 10^{11}$
MeV$^{4}$ (dotted-dashed curve). We have taken $B^{1/4}=200$ MeV
, $G_4\sim10^{-44}$ MeV$^{-2}$ and $\ell_{_{\rm DGP}}\sim10^{-38}$ MeV$^{-1}$.}
\end{figure}

It is well known in brane-world models that a quadratic term in
the energy density appears in the Friedmann equation and that in
the DGP scenario such a term can be obtained by taking the
negative branch ($\epsilon=-1$) when  $H\ll\ell_{_{\rm
DGP}}^{-1}$, describing the effects of the extra dimensions. Thus,
at the beginning the quadratic term dominates in the evolution of
the universe so that the evolution of the hadronic fraction during
the phase transition is given by 
\begin{equation}\label{eq4.15}
\frac{dh}{d\tau}=-\frac{(1+rh)}{r}[8\pi G_{4}\rho _{_{Q}}(1+nh)+\lambda]
,
\end{equation}
with the general solution given by
\begin{equation}\label{eq4.16}
h(\tau)=\frac{\exp[(8\pi G_{4}\rho
_{_{Q}}+\lambda)\tau+\frac{8\pi G_{4}n}{r}C']-(8\pi G_{4}\rho
_{_{Q}}+\lambda)\exp\left[\frac{8\pi G_{4}\rho
_{_{Q}}n}{r}\tau+rC'(8\pi G_{4}\rho
_{_{Q}}+\lambda )\right]}{n8\pi G_{4}\rho
_{_{Q}}\exp\left[\frac{8\pi G_{4}\rho
_{_{Q}}n}{r}\tau+rC'(8\pi G_{4}\rho
_{_{Q}}+\lambda )\right]-r\exp[(8\pi G_{4}\rho
_{_{Q}}+\lambda)\tau+\frac{8\pi G_{4}n}{r}C']}
.
\end{equation}
The integration constant $C'$ can be obtained by relating it to the
$h\left(\tau_{c}\right)=0$  at the
time $\tau_{0}$ by using the relation
\begin{equation}\label{eq4.17}
C'=\frac{8\pi G_{4}\rho
_{_{Q}}n\tau_{0}+r\ln[\frac{e^{(8\pi G_{4}\rho
_{_{Q}}+\lambda)\tau_{0}}}{8\pi G_{4}\rho
_{_{Q}}+\lambda}]}{r(8\pi G_{4}\rho
_{_{Q}}n-8\pi G_{4}\rho
_{_{Q}}r-\lambda r)}.
\end{equation}
From  equation (\ref{eq4.13}), the evolution
of the scale factor on the brane, associated with the hadronic
fraction during the phase transition 
dominated by the evolution of the extra dimension of the cosmological fluid, can be obtained.
We note that the end of the phase transition era corresponds to
$h(\tau)=1$, thus one obtains the time $\tau_{h}$ in which the phase
transition ends.
\subsection{Pure hadronic era}
At the end of the phase transition the scale factor of the
universe has the value
\begin{equation}
a_{_{0}}\left( \tau_{h}\right)
=a_{_{0}}\left( \tau_{c}\right) \left( r+1\right) ^{\frac{-1}{3}}.
\end{equation}
Also, the energy density of the pure hadronic matter after the phase
transition is $\rho _{h}=3p_{h}=3a_{\pi }T^{4}$. The conservation
equation on the brane (\ref {eq2.9}) gives
$a_{_{0}}(T)=a_{_{0}}\left( t_{h}\right) T_{c}/T$. The temperature
dependence of the DGP brane universe in the hadronic phase is
governed by the equation
\begin{equation}\label{eq4.19}
\frac{dT}{d\tau}=-T\left(\frac{\epsilon}{2\ell_{_{\rm DGP}}}+\sqrt{\frac{1}{4\ell_{_{\rm DGP}}^2}+\lambda+
8\pi G_4 a_\pi T^{4}}\right).
\end{equation}
Variation of temperature of the hadronic fluid filled $\lambda$DGP brane
universe as a function of $\tau$ for different values of
$\lambda $ is presented in figure 3.
\begin{figure}
\begin{center}
\epsfig{figure=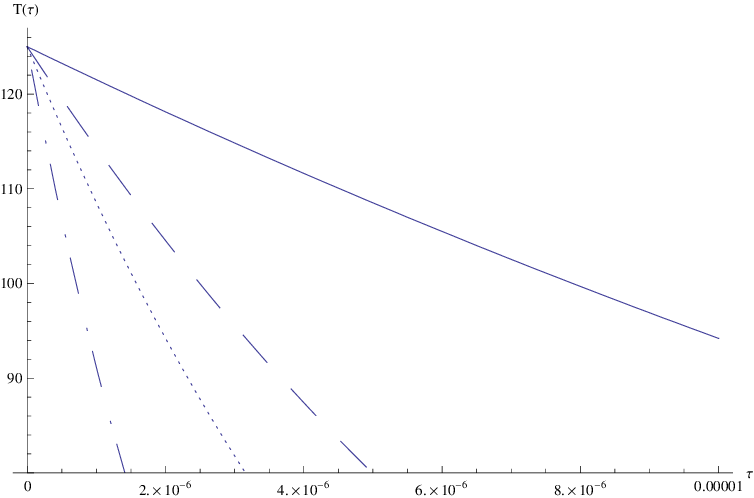,width=8cm}
\end{center}
\caption{\footnotesize  The behavior of $T(\tau)$ as a function of
time ($\tau$) for the positive branch ($\epsilon=1$) and different
values of $\lambda$: $\lambda=8\times10^{8}$
MeV$^{4}$ (solid curve), $\lambda=8\times 10^{9}$
MeV$^{4}$ (dashed curve), $\lambda=2\times 10^{10}$
MeV$^{4}$ (dotted curve) and $\lambda=1\times 10^{11}$
MeV$^{4}$ (dotted-dashed curve). We have taken $B^{1/4}=200$ MeV
, $G_4\sim10^{-44}$ MeV$^{-2}$ and $\ell_{_{\rm DGP}}\sim10^{-38}$ MeV$^{-1}$.}
\end{figure}
\section{Lattice QCD equation of state}

QCD phase transition is a fundamental concept in particle
physics having become increasingly relevant to any study
dealing with the underlying mechanisms responsible
for the evolution of the early universe. In such a
scenario, a soup of quarks and gluons interact and undergo
a crossover transition to form hadrons. It is therefore useful
to have a brief review of the basic notions before using the results obtained from such a phase transition and
apply them to the study of the universe at early times within
the context of DGP brane scenario.
Lattice QCD is a new approach which allows one to systematically
study the non-perturbative regime of the QCD
equation of state. Employing supercomputers, the QCD equation
of state was calculated on the lattice in \cite{66} with two
light quarks and a heavier strange quark on a $(N_t = 6)×32^{3}$
size lattice. The quark masses have been chosen to be close
to their physical values, i.e. the pion mass is about $220 MeV$.
For further details we refer the reader to \cite{66}. However, we
remark that the equation of state was calculated at a temporal
extent of the lattice $N_t = 6$ and for $N_t = 6$, sizable lattice
cut-off effects are still present \cite{67}. The data for energy
density $\rho(T )$, pressure $p(T)$ and trace anomaly $\rho -3p$ and
entropy $s$ are from  \cite{66}. The analysis in this section
uses these data. Besides the strange quark,
one can also include the effect of the charm quark as well
as photons and leptons on the equation of state. These have
important cosmological contributions as was shown in \cite{68}.
Recent information on lattice QCD at high temperature can be found in
\cite{69--71}.
During the high temperature regime we see, as expected,
radiation like behavior in the region at and below the critical
temperature $T_c (\approx200$ MeV) of the de-confinement transition
where the behavior changes drastically. This change in behavior
will also be relevant for cosmological observables as
we will see below. For high temperatures between
$2.82 (100$ MeV) and $7.19 (100$ MeV) one can fit the data to
a simple equation of state of the form
\begin{eqnarray}\label{eqq1}
\rho_{_{High}}(T) \approx \alpha T^{ 4},\\\nonumber
p_{_{High}}(T ) \approx \sigma T ^{4}.
\end{eqnarray}
One can then find $\alpha = 14.9702 \pm 009997$ and $\sigma = 4.99115 \pm 004474$ using a least squares fit \cite{66}.
While for times before the phase transition the lattice data
match the radiation behavior very well, for times corresponding
to temperatures above $T_c$ the behavior of the lattice
data changes towards that of the matter dominated phase. We
remark that lattice studies show that  QCD phase transition
is actually a crossover transition.

\section{ Hadronic resonance gas model}
Besides lattice QCD there are other approaches to the
low temperature equation of state. In the framework of the
Hadronic Resonance Gas model (HRG), QCD in the confinement
phase is treated as an non-interacting gas of fermions
and bosons \cite{72}. The fermions and bosons in this
model are the hadronic resonances of QCD, namely mesons
and baryons. The idea of the HRG model is to implicitly
account for the strong interaction in the confinement phase
by looking at the hadronic resonances only, since these are
basically the relevant degrees of freedom in that phase. The
HRG model is expected to give a good description of thermodynamic
quantities in the transition region from high
to low temperatures \cite{76}. The HRG result for the trace
anomaly can also be parameterized by the simple form \cite{78}
\begin{eqnarray}\label{eqq2}
\frac{\Theta(T)}{T^{4}}
\equiv
\frac{\rho - 3p}{T^4}= a_1T +a_2T^{ 3} +a_3T^{ 4} + a_4T ^{10},
\end{eqnarray}
with $a_1 = 4.654 GeV^{-1}$, $a_2 = -879 GeV^{-3},$ $a_3 = 8081 GeV^{-4} $ and $a_4 =-7039000 GeV^{-10}$.

In lattice QCD, the calculation of the pressure, energy
density and entropy density usually proceeds through the
calculation of the trace anomaly $\Theta(T ) = \rho(T ) -3p(T )$.
Using the well known thermodynamic identity, the pressure difference at
temperatures $T$ and $T_{low}$ can be expressed as the integral of
the trace anomaly
\begin{eqnarray}\label{eqq3}
\frac{p(T )}{T_4}-\frac{p(T_{low})}{T^{4}_{low}}=\int^{T}_{T_{low}}\frac{dT'}{T^{5}\Theta(T')}.
\end{eqnarray}
By choosing the lower integration limit sufficiently small,
$p(T_{_{low}})$ can be neglected due to the exponential suppression and the energy density $\rho(T ) = \Theta(T ) + 3p(T )$ and
entropy density $s(T ) = (\rho+ p)/T$ can be calculated.
This procedure is known as the integral method \cite{79}. Using
(\ref{eqq2}) and (\ref{eqq3}), we obtain
\begin{eqnarray}\label{eqq4}
\rho_{_{Low}}(T ) = \eta T^4 +4a_1T^{5} + 2a_2T^7 +\frac{7a_3}{4}T^8 +\frac{13a_4}{10}T^{14},
\end{eqnarray}
where $\eta =-0.112.$ The trace anomaly plays a central role
in lattice determination of the equation of state. The equation
of state is obtained by integrating the parameterizations
given in (\ref{eqq2}) over temperature as shown in (\ref{eqq3}).

\section{ DGP  brane universe and QCD phase transition}
\subsection{High temperature regime}
Let us first consider the era before the phase transition at
high temperature where the universe is in the quark phase.
Using the conservation equation of matter together with the
equation of state of quark matter (\ref{eqq1}), one gets the following
relation for the Hubble parameter
\begin{equation}\label{eqq5}
H = \frac{\dot{a}}{a} =-\frac{4\alpha}{3(\alpha+\sigma)}\frac{\dot{T}}{T},
\end{equation}
whence one can solve for the scale factor
\begin{equation}\label{eqq6}
a(T ) = cT^{\frac{-4\alpha}{3(\alpha+\sigma)}},
\end{equation}
where c is a constant of integration.

One may now proceed to obtain an expression describing
the behavior of temperature of the DGP brane universe
as a function of time i the quark phase. Using (\ref{eqq1}) and (\ref{eq2.10}), one finds a differential equation for the temperature as follows
\begin{equation}\label{eqq7}
\frac{dT}{dt}=-\frac{3(\alpha+\sigma)T}{4\alpha}\left[{\epsilon\over 2\ell_{_{\rm DGP}}}+\sqrt{\frac{1}{4\ell_{_{\rm DGP}}^2}+{8\pi G_4\over 3}\rho_{_{High}}+\lambda}\right],
\end{equation}
where we have set the constant $c$ to  unity.
The lower temperature limit in our definition comes from the fact that
for $T <170 $MeV there is agreement with the HRG equation
of state.
Equation (\ref{eqq7}) can be solved numerically and the result is
plotted in Fig. 4, which shows the behavior of  temperature
of the universe in the quark phase as a function of cosmic
time in DGP cosmology for different values of the cosmological constant on the brane, $\lambda$, in the interval $282$ MeV$<T <719 $MeV in
high temperature regime. It is seen that as the time evolves
the universe becomes cooler. In Fig. 4 it is seen that the cosmological constant on the brane
enhances the cooling process because a larger slope is
observed for a large coupling, $\lambda$.
\begin{figure}
\begin{center}
\epsfig{figure=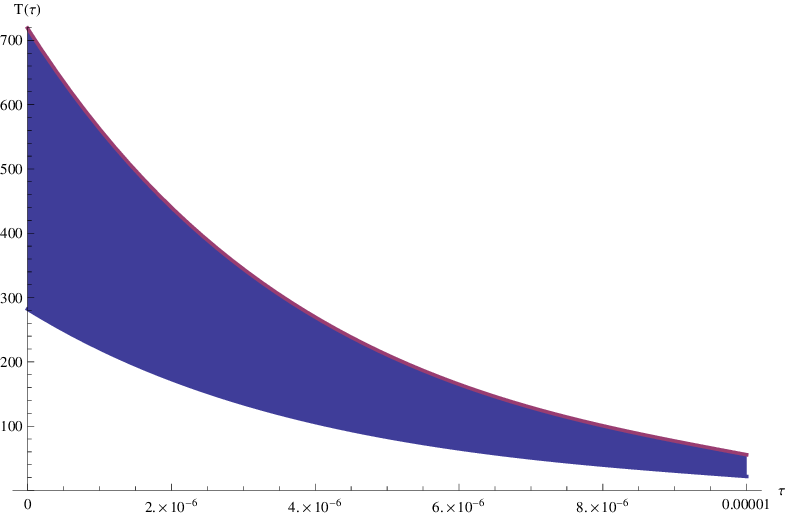,width=6cm}\hspace{5mm}
\epsfig{figure=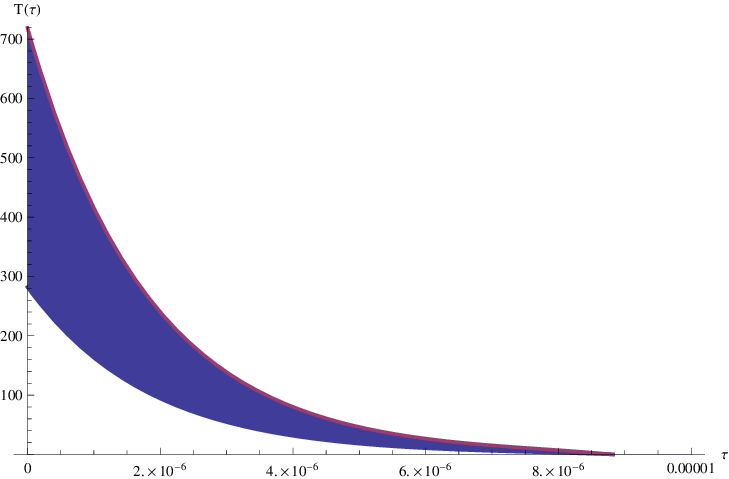,width=6cm}
\end{center}
\caption{\footnotesize The behavior of $T (\tau)$ in interval $282$ MeV $< T < 719$ MeV
as a function of $\tau$ for the positive branch ($\epsilon=1$) and the values of $\lambda=6\times10^{10}$ MeV$^{4}$ (left) and $\lambda=3\times10^{11}$ MeV$^{4}$ (right). We have taken $G_4\sim10^{-44}$ MeV$^{-2}$ and $\ell_{_{\rm DGP}}\sim10^{-38}$ MeV$^{-1}$
.}
\end{figure}

\subsection{Low temperature regime}
Let us now consider the era before phase transition
at low temperature where the universe is in the confinement
phase and treated as a non-interacting gas of fermions and
bosons \cite{72}. Using the conservation equation of matter
together with equation of state (\ref{eqq4}), one gets the
following relation for the Hubble parameter
\begin{equation}\label{eqq8}
H =\frac{\dot{a}}{a} =-\frac{12\eta T^3 +20a_1T^4 +A(T)}{3[4\eta T^4 + 5a_1T^5 +B(T)]}\dot{T},
\end{equation}
where
\begin{eqnarray}\label{eqq9}
A(T) = 14a_2T^6 +14a_3T^7 +\frac{91}{5}T^{13},\\\nonumber
B(T) =73a_2T^7 +2a_3T^8 +75a_4T^{14}.
\end{eqnarray}
We may now solve for the scale factor and find
\begin{equation}\label{eqq10}
a(T ) =\frac{c}{T (75a_1T +35a_2T^3 +30a_3T^4 +21T^{10} +60\eta)^{1/3}} ,
\end{equation}
where $c$ is a constant of integration.
We can obtain an expression describing the behavior of
temperature of the DGP brane universe as a function of
time in the quark phase.
Upon using (\ref{eqq4}) and (\ref{eq2.10}), one finds a differential equation for  temperature as follows
\begin{eqnarray}\label{eqq11}
\frac{dT}{dt}=-\frac{3[4\eta T^4 + 5a_1T^5 +B(T)]T}{12\eta T^3 +20a_1T^4 +A(T)}\left[{\epsilon\over 2\ell_{_{\rm DGP}}}+\sqrt{\frac{1}{4\ell_{_{\rm DGP}}^2}+{8\pi G_4\over 3}\rho_{_{Low}}+\lambda}\right]
\end{eqnarray}
where we have set the constant $c$ to  unity.
Equation (\ref{eqq11}) may now be solved numerically and the result is plotted
in Fig. 5, which shows the behavior of  temperature of
the universe in the quark phase as a function of the cosmic time
for different values of $\lambda$, in the interval $80$ MeV $< T < 180$ MeV at
the low temperature regime. From Fig. 5 it can be seen that in
the context of the HRG model,  $\lambda$ enhances the cooling
process. It seems in our model the phase transition
depends on $\lambda$ and  can be interpreted as
a running coupling constant for the phase transition.
\begin{figure}
\begin{center}
\epsfig{figure=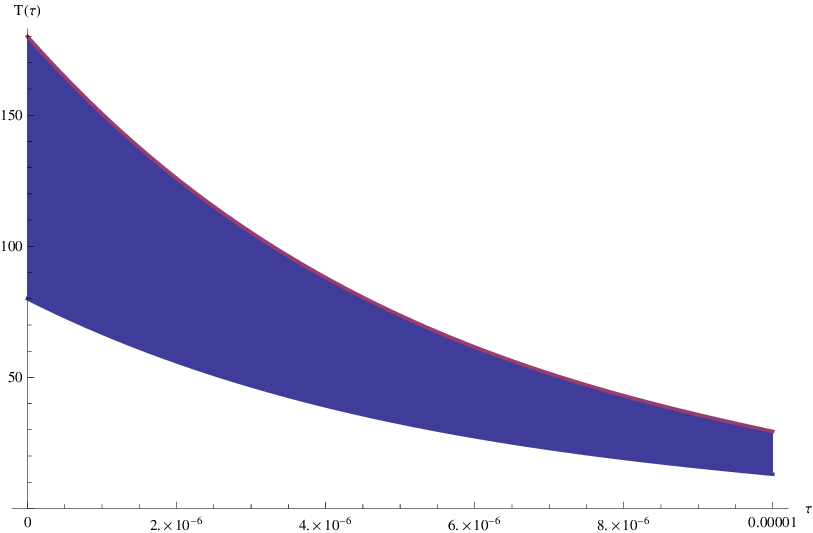,width=6cm}\hspace{5mm}
\epsfig{figure=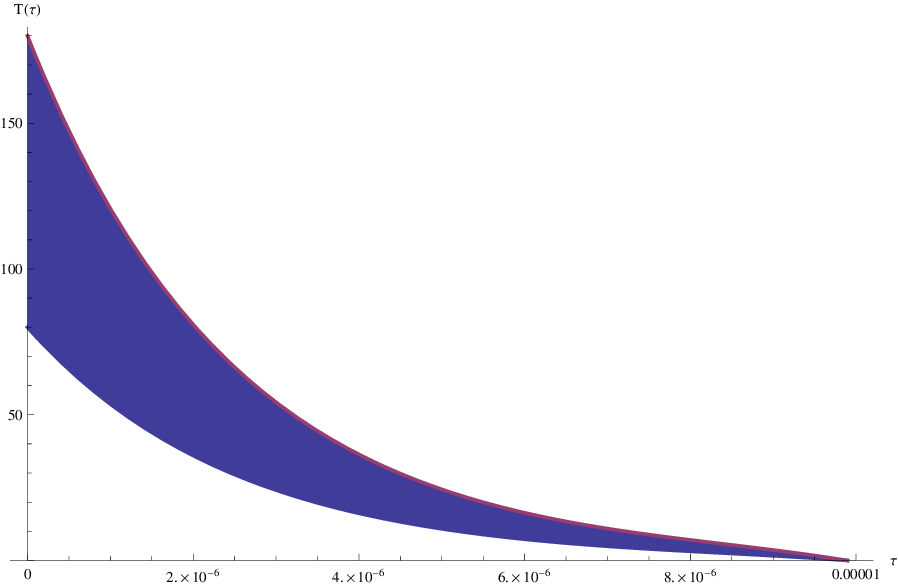,width=6cm}
\end{center}
\caption{\footnotesize The behavior of $T (\tau)$ in interval $80$ MeV $< T < 180$ MeV
as a function of $\tau$ for the positive branch ($\epsilon=1$) and the values of $\lambda=6\times10^{11}$ MeV$^{4}$ (left) and $\lambda=3\times10^{12}$ MeV$^{4}$
(right). We have taken $G_4\sim10^{-44}$ MeV$^{-2}$ and $\ell_{_{\rm DGP}}\sim10^{-38}$ MeV$^{-1}$.}
\end{figure}

\section{Conclusion}
In this paper, we have discussed the quark–-hadron phase transition
in a DGP brane cosmological scenario where our universe is a
three-brane embedded in a $5$-dimensional bulk  with an
intrinsic Ricci curvature scalar and a cosmological constant, 
within an effective model of QCD. We have studied  the evolution of
the physical quantities, relevant to the physical description of the
early universe; the energy density, temperature and scale factor,
before, during, and after the phase transition. In particular, due
to the high energy density in the early universe, we have studied in
detail the specific case where the terms
linearly proportional to the energy density can be neglected relative to the
quadratic terms in the Friedmann equation when the negative branch
is considered.

Finally, in section 5 we have treated the time evolution of temperature and scale factor
which are relevant to the physical portrayal of the universe at
early times in the crossover transition for high and low temperatures.
We have shown that for different values of $\lambda$ the phase transition occurs and causes
the effective temperature of the quark–gluon
plasma and of the hadronic fluid to decrease. Comparing Figs. 4 and 5
we see that the slope of temperature, $T$ , for the above mentioned two
models is different during the crossover transition. It is seen that the slope of $T$  at low temperature for the HRG
model is steeper than for the high temperature regime. In the smooth crossover regime where lattice
QCD is used to investigate the high temperature behavior,
we see that the slope is smooth relative to first order
phase transition while at lower temperatures where HRG is
used, the slope is steep compared to first order phase transition.
Taking into account the energy range in which the
calculations are done, one might conclude that these two approaches
to the quark–-hadron transition in the early universe
do not predict fundamentally different ways of the evolution
of the early universe.


\end{document}